\documentclass[aps,prl,showpacs,twocolumn,superscriptaddress]{revtex4-1}
\usepackage{graphicx}
\usepackage{amsmath}
\usepackage{bbold}
\usepackage{ipa}
\usepackage{makeidx}
\usepackage[usenames,dvipsnames,svgnames]{xcolor}

\def\im{\mathrm{Im}}

\def\re{\mathrm{Re}}

\def\simge{\mathrel{%
    \rlap{\raise 0.511ex \hbox{$>$}}{\lower 0.511ex \hbox{$\sim$}}}}
\def\simle{\mathrel{
    \rlap{\raise 0.511ex \hbox{$<$}}{\lower 0.511ex \hbox{$\sim$}}}}

\newcommand \be{\begin{eqnarray}}
\newcommand \ee{\end{eqnarray}}

\def\XXint#1#2#3{{\setbox0=\hbox{$#1{#2#3}{\int}$}
\vcenter{\hbox{$#2#3$}}\kern-.5\wd0}}

\begin{document}

\title{Dysonian dynamics of the Ginibre ensemble}

\author{Zdzislaw Burda}
\email{zdzislaw.burda@uj.edu.pl} \affiliation{M. Smoluchowski Institute
of Physics and Mark Kac Center for Complex Systems Research,
Jagiellonian University, PL--30--059 Cracow, Poland}

\author{Jacek Grela} \email{jacek.grela@uj.edu.pl} \affiliation{M. Smoluchowski Institute
of Physics and Mark Kac Center for Complex Systems Research,
Jagiellonian University, PL--30--059 Cracow, Poland}

\author{Maciej A. Nowak}
\email{nowak@th.if.uj.edu.pl} \affiliation{M. Smoluchowski Institute
of Physics and Mark Kac Center for Complex Systems Research,
Jagiellonian University, PL--30--059 Cracow, Poland}

\author{Wojciech Tarnowski} \email{wojciech.tarnowski@uj.edu.pl} \affiliation{M. Smoluchowski Institute
of Physics and Mark Kac Center for Complex Systems Research,
Jagiellonian University, PL--30--059 Cracow, Poland}

\author{Piotr Warcho\l{}} \email{piotr.warchol@uj.edu.pl} \affiliation{M. Smoluchowski Institute
of Physics and Mark Kac Center for Complex Systems Research,
Jagiellonian University, PL--30--059 Cracow, Poland}
\date{\today}

\begin{abstract} 
We study the time evolution of Ginibre matrices whose elements undergo Brownian motion.
The non-Hermitian character of the Ginibre ensemble binds the dynamics of eigenvalues to the evolution of eigenvectors in a non-trivial way,
leading to a system of coupled nonlinear equations resembling those 
for turbulent systems. We formulate  a mathematical framework allowing  simultaneous  description of  
the flow of eigenvalues and eigenvectors, and we unravel a hidden dynamics 
as a function of new complex variable, which in the standard description is treated as a regulator only.  We solve the evolution equations for large matrices and demonstrate that the non-analytic behavior of the Green's functions is associated with a shock wave stemming from a Burgers-like equation describing correlations of eigenvectors. We conjecture that the hidden dynamics, that we observe for the Ginibre ensemble, is a general feature of non-Hermitian random matrix models and is relevant to related physical applications.\end{abstract}

\maketitle
Today, half a century after the pioneering work of Ginibre \cite{GG}, random matrices with complex spectra are no longer only of academic interest. They play a role in quantum information processing \cite{KAROLZ}, in QCD in problems with a finite chemical potential \cite{CHEMICALQCD}, in financial engineering when lagged correlations are discussed \cite{BT} and in the research on neural networks \cite{NEURAL}, to name just a few applications. Eigenvalues themselves, however, are not of sole interest in the case of non-Hermitian random matrix ensembles. The statistical properties of eigenvectors are equally significant \cite{CHALKERMEHLIG}, in particular, in problems concerning scattering in open chaotic cavities or random lasing \cite{PETER1, CHSCATTER, HVH}. There, the so called Petermann factor \cite{PETERMANN}, a quantity describing correlations between right and left eigenvectors, modifies the quantum-limited line-width of a laser.

On the other hand, the original Dyson's idea of Brownian walk of real eigenvalues \cite{DYSON} interacting with a two-dimensional Coulombic force still leads to novel insights. Examples include the study of determinantal processes \cite{MAJUMDAR,KATORI, SCHEHR},  Loewner diffusion \cite{LOEWNER} or the fluctuations of non-intersecting interfaces in thermal
equilibrium \cite{NADAL}. The concept of stochastic evolution of matrices has been recently exploited by several authors~\cite{BN1,BNW1,NEUBERGER,EWA}. In particular, it was shown that the derivatives of the logarithms of characteristic determinants of diffusing GUE (Gaussian Unitary Ensemble), LUE (Laguerre Unitary Ensemble or Wishart Ensemble) and CUE (Circular Unitary Ensemble) obey Burgers-like nonlinear equations, where the role of viscosity is played by the inverse of the matrix size.  For infinite dimensions of the matrix, these equations correspond to the inviscid regime and describe an evolution of the associated resolvents. Due to nonlinearity, they develop singularities (shock waves), whose positions correspond to the endpoints of the spectra. For matrices of finite size, the expansion around the shock wave solution of the initial viscid Burgers equation leads to a universal  scaling of characteristic polynomials (and of the inverse characteristic polynomials as well), resulting in well known universal oscillatory behavior of the Airy, Bessel or Pearcey type. This approach has prompted, in particular, new perception of  weak/strong coupling transition  in multicolor Yang-Mills theory \cite{NEUBERGERPRL,BN0} and of the spontaneous breakdown of chiral symmetry in Euclidean QCD \cite{BNWWISHART2}.

In this letter, we unveil the intertwined evolution of eigenvalues and eigenvectors of stochastically evolving non-Hermitian matrices. To this end, we apply   Dyson's idea  to study diffusing Gaussian matrices for the case of  the Ginibre Ensemble (GE). The central object of the paper is a generalized averaged characteristic polynomial. Its logarithmic derivatives, which contain the information about both the eigenvalues and eigenvectors of the evolving matrix, fulfill a system of Burgers-like partial differential equations. We solve them to recover the spectral density, the Petermann factor encoding the correlations of eigenvectors and universal microscopic scaling at the edge of the support of the eigenvalues.  

At first glance one would not expect any similarities between the GUE and the GE, even in the large $N$ (matrix' size) limit.  In the case of GUE, spectra are real, endpoints of the spectra exhibit square root behavior and the eigenvectors decouple from the eigenvalues. In the case of GE, spectra are complex, eigenvalues form a uniform disc with a vertical cliff at the boundary and finally, left and right eigenvectors are correlated \cite{CHALKERMEHLIG} on the support of eigenvalues.  Nonetheless, the Vandermonde determinant is present in the joint probability distribution of eigenvalues for both ensembles and this 
leads to a two-dimensional electrostatic Dyson's picture which
underlies calculations of the spectral distribution in the large $N$ limit. 
Consequently, the standard procedure for non-hermitain ensembles relies on defining the electrostatic potential
\be
V(z)=\lim_{\epsilon \rightarrow 0} \lim_{N \rightarrow \infty} \frac{1}{N} \left\langle {\rm Tr} \ln [|z-X|^2 +\epsilon^2] \right\rangle \ ,
\label{Coulomb}
\ee
calculating the "electric field" as its gradient, 
$G=\partial_z V$, and recovering the spectral function from the Gauss law 
$\rho =\frac{1}{\pi}\partial_{\bar{z}}G = \frac{1}{\pi} \partial_{z\bar{z}} V$.  
We use a short-hand notation defined by:
$|z-X|^2 + \epsilon^2 = (z\mathbb{1}_N -X)(\bar{z}\mathbb{1}_N-X^{\dagger}) + \epsilon^2 \mathbb{1}_N$, where $\mathbb{1}_N$ is the $N$-dimensional identity matrix. 
$\epsilon$ is an infinitesimal regulator and it is crucial that the limit $N\rightarrow \infty$ is taken first. If one took the limits
in an opposite order, one would obtain a trivial result.
Moreover, in the case of the Ginibre ensemble, $\left\langle \det (z-X) \right\rangle =z^N$. The standard relation between zeros of the characteristic polynomials and 
poles of the Green's function, known from considerations of hermitian ensembles, 
would therefore be lost. 

The idea is to define the following object
\be
D(z,w,\tau)=\left\langle \det(Q -H) \right\rangle_{\tau} \nonumber \\
 =\left\langle \det \left( |z-X|^2 +|w|^2 \right) \right\rangle_{\tau} ,
\label{basic}
\ee
where
\be
Q= \left( \begin{array}{cc}
z & -\bar{w} \\ w & \bar{z} \end{array} \right) ,  \,\,\,\,\,\,\,\,H=\left( \begin{array}{cc}
X & 0 \\ 0 & X^{\dagger} \end{array} \right).
\ee
and to study its evolution in the space of $Q$, or more precisely
in the complex plane $w$, "perpendicular" to the basic complex plane $z$. 
In other words, the regulator $i \epsilon$, which is usually treated
as an infinitesimally small real variable, is promoted
to a genuine complex variable $w$. As we shall see, the dynamics of
$D(z,w,\tau)$ hidden in $w$ captures the evolution of eigenvectors 
and eigenvalues of the Ginibre matrix whose elements undergo Brownian motion. It is worth mentioning that block matrices such as $H$ and arguments $Q$ naturally appear in non-Hermitian random matrix models, e.g. in the generalized Green's function technique \cite{JANIKNOWAK,JAROSZNOWAK}, in hermitization methods \cite{GIRKO,FEINBERGZEE,CHALKERWANG} or in the derivation
of the multiplication law for non-Hermitian random matrices \cite{BJN}. 

In our notation, the meaning of the averages
$\left\langle \ldots \right\rangle_{\tau}$ like this
in (\ref{basic}) is
$\left\langle F(X) \right\rangle_{\tau} = 
\int DX P(X,\tau|X_0,0) F(X)$, where
$DX = \sum_{ab} d x_{ab} d y_{ab}$ is a flat measure
over the real and imaginary parts of matrix elements, 
$X_{ab} = x_{ab} + i y_{ab}$, and $P(X,\tau|X_0,0)$ is the
probability that the matrix will change from its 
initial state $X_0$ at $\tau=0$ to $X$ at time $\tau$.
For a free random walk with independent increments 
$\left\langle \delta X_{ab} \right\rangle_\tau=0$ 
and $\left\langle \delta X_{ab} \delta \bar{X}_{cd} \right\rangle_\tau
= \frac{\delta \tau}{N} \delta_{ac}\delta_{bd}$, the evolution of
$P(X,\tau|X_0,0)$ is governed by the diffusion equation 
\be
\partial_{\tau}P(X,\tau|X_0,0)=\frac{1}{N} \partial_{XX^{\dagger}} P(X,\tau|X_0,0),
\label{diffusion}
\ee
where $\partial_{XX^{\dagger}}$ is the standard $2N^2$-dimensional Laplacian 
$\partial_{XX^{\dagger}} = \sum_{ab} \left(\partial^2_{x_{ab}} + \partial^2_{y_{ab}}\right)$.
The announced  dynamics of the Ginibre ensemble is hidden in equation
\be
\partial_{\tau} D(z,w,\tau) = \frac{1}{N} \partial_{w\bar{w}} D(z,w,\tau)
\label{main}
\ee
that is central to this paper.  The derivation will be presented elsewhere, but  we shortly sketch below  the main
steps. The determinant in (\ref{basic}) can be represented as a Berezin 
integral $\int \exp\left[\theta^T(Q-H)\eta\right] \,
d\theta\,d\eta = \det (Q-H)$ where $\theta$ and $\eta$ are independent 
vectors of Grassmann variables. Both sides of eq.
(\ref{diffusion}) can be then multiplied by this integral and 
integrated over $DX$. After some manipulations, 
like changing the order of integration and integrating by parts,
one arrives at (\ref{main}). 

A few comments are in order. First, it is easy to see that $D(z,w,\tau)$ depends
on $w$ only through its modulus $r=|w|$. Moreover, the simplest initial condition corresponds
to $X_0=0$. In this case $D_0(z,w)=D(z,w,0)=(|z|^2 + |w|^2)^N$. Finally, it is instructive
to compare $D_0$ to the initial determinant for another matrix, one that
also has all eigenvalues equal to zero, for instance a strictly upper
triangular matrix. As an example consider a matrix $X'_0$, with all elements 
equal to zero except a single off-diagonal element that is equal to one. 
For $X'_0$, the initial value of the determinant is 
$D'_0(z,w)= (|z|^2 + |w|^2)^N \left( 1 + |w|^2/(|z|^2+|w|^2)^2\right)$ and,
as we can see, for $|w|\ne 0$, it differs from $D_0(z,w)$. This simple example shows
that the dependence of the determinant on $w$ indeed encodes far more 
information on the underlying matrix than just its eigenvalues. Such information, as we shall see below, is very valuable.
 
We proceed by defining two convenient functions
$v=v(z,r,\tau)$ and $g=g(z,r,\tau)$: 
\be
v \equiv \frac{1}{2N} \partial_r \ln D, \label{mainv}\\
g \equiv \frac{1}{N} \partial_z \ln D \label{maing},
\ee
which will turned out to be  closely related to the eigenvector 
correlator and the Green's function known from the standard 
treatment of the Ginibre ensemble. These functions  are not independent, since
by construction $\partial_z v = \frac{1}{2} \partial_r g$; in particular
$g = 2 \int dr \partial_z v$.
The diffusion equation (\ref{main}) is mapped via   
(\ref{mainv}), which basically is the inverse Cole-Hopf transformation 
\cite{COLEHOPF}, onto a Burgers-like equation 
\be
\partial_{\tau}v &=& v\partial_r v + \frac{1}{N} \left(\Delta_r - \frac{1}{4r^2}\right) v ,
\label{twoBurgers}
\ee 
where $\Delta_r=\frac{1}{4}(\partial_{rr} +\frac{1}{r}\partial_r)$ is the radial 
part of the two-dimensional Laplacian. This equation
is exact for any $N$. The $1/N$ factor is a viscosity-like parameter. 
In the inviscid limit ($N \rightarrow \infty$), (\ref{twoBurgers}) reduces to
\be
\partial_{\tau}v&=v\partial_r v,
\label{Euler}
\ee
known in hydrodynamics as the Euler equation and solved by  the method of characteristics.
The curves along which the solution is constant are namely given by 
\be
r=\xi-v_0(\xi)\tau,
\label{charact}
\ee
and labeled with $\xi$. $v_0$ plays the role of velocity of the front-wave.
We therefore have
\be
v=v_0(r+\tau v).
\ee
For the initial condition $X_0=0$,  
corresponding to $v_0(r) =r/(z\bar{z}+r^2)$, we obtain a cubic algebraic equation for $v$. 
Its solution gives in turn the (radial) dependence of $v$ on $r=|w|\ge 0$. 
If one takes a cross-section of the whole solution along the 
real axis, $\im \; w=0$ and $\re \; w = \mu$ (or any other straight
line going through the origin of the $w$-complex plane),
one can see that in fact the solution consists of two symmetric branches $v(\mu)=v(-\mu)$ due to the rotational symmetry of the problem
in the complex plane. In other words, the solution is represented by the pair of Cardano equations:
\be
v \; \left(z\bar{z} +(\pm \mu + \tau v)^2\right)  = \pm \mu + \tau v,
\label{Cardano2}
\ee
since $\mu$, as opposed to $r$, may be positive or negative. The mapping between $r$ and $\xi$ breaks down when,  at some  positions $\mu=\pm r_*$, derivative becomes singular ($d\xi/dr_*=\infty$), as visualized on the left inlet at Fig.~1. The set of singular points defines  the caustics (sometimes called pre-shocks). 
Physically, the singularity comes from the 
fact that the velocity of the flow is position-dependent, which makes the solution, for a given $|z|$, non-unique after a certain time $\tau$. Between the two symmetric caustics 
(which actually form a cone-like surface when viewed from the whole $w$-complex-plane)
a shock is formed at $\mu=0$ for
$\tau \ge |z|^2$.
Although the shock formation involves the whole $(w,z)$ space, as depicted in Fig.~1, its dynamics is remarkably confined to the region of $r=|w|\to 0$, close to the $z$-plane, which is the basic complex plane in our considerations. As was already mentioned,
in this region $r$ plays the role of the regulator $\epsilon$ in the
formula (\ref{Coulomb}).  
In this limit the explicit solution of (\ref{Cardano2}) reads 
\be
 v^2=(\tau-|z|^2)/\tau^2 \quad {\rm and} \quad v=0, \quad \text{as r} \to 0.
\label{circle}
\ee
The quantity $v^2$ has an explicit interpretation \cite{NOWAKNOER} in the large $N$ limit, namely
\be
v^2 = \frac{\pi}{N^2} \left< \sum_i A_{ii}\delta^2(z-\lambda_i) \right>,
\ee
where $A_{ij}=\left <L_i|L_j\right >\left <R_j|R_i \right>$, i.e. $v^2$ is a correlator between the bi-orthogonal sets of left $|L_i\rangle$ and right $|R_i\rangle$
eigenvectors of the non-hermitian matrix $X$ to eigenvalues $\lambda_i$, originally introduced in \cite{CHALKERMEHLIG}. Modulo normalization, this correlator is also known from chaotic scattering theory as the Petermann factor \cite{PETER1}:
\begin{align}
	K(z,\tau) = \frac{N v^2}{\pi \rho}
\end{align} 
(where $\rho$ is the spectral density calculated later). Figure~2 shows the time dependence of Petermann factor for several values of $|z|$. 
The correlator vanishes
outside the critical shock line, where, as we know from the standard approach,
the Green's function is analytic, and it is non-zero inside it, where the Green's function is non-analytic. The edge of the shock line lines up with the contour of the eigenvalue density support. To summarize, the quaternion shock wave dynamics (\ref{circle}) reproduces the result of \cite{CHALKERMEHLIG}. 
\begin{figure}[htbp]
\centering
\includegraphics[width=0.49\textwidth]{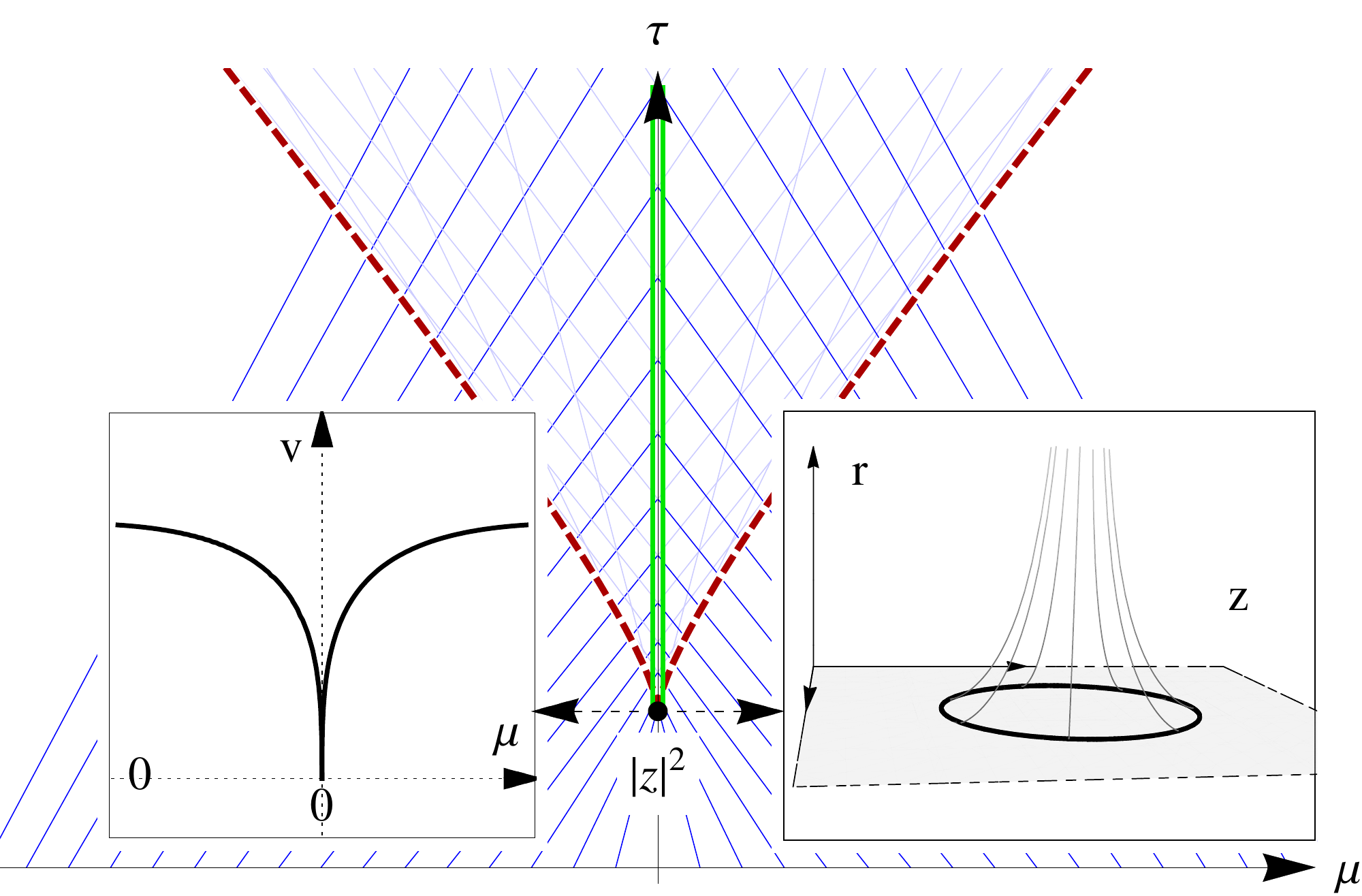}
\caption{The main figure shows, for a given $|z|$, the characteristics (straight lines) and caustics (dashed  lines). Inside the later a shock is developed 
(double vertical line). Left  inlet shows
 the solution of eq. (\ref{Cardano2})
 at ($\tau = |z|^2$).
Right  inlet shows the caustics  mapped to the $(r=|w|,z)$ hyperplane at the same moment of time. The section $r=0$ yields the circle $|z|^2=\tau$, bounding the domain of eigenvalues and eigenvectors correlations for the GE. }
\label{plot}
\end{figure}
\begin{figure}[htbp]
\centering
\includegraphics[width=0.49\textwidth]{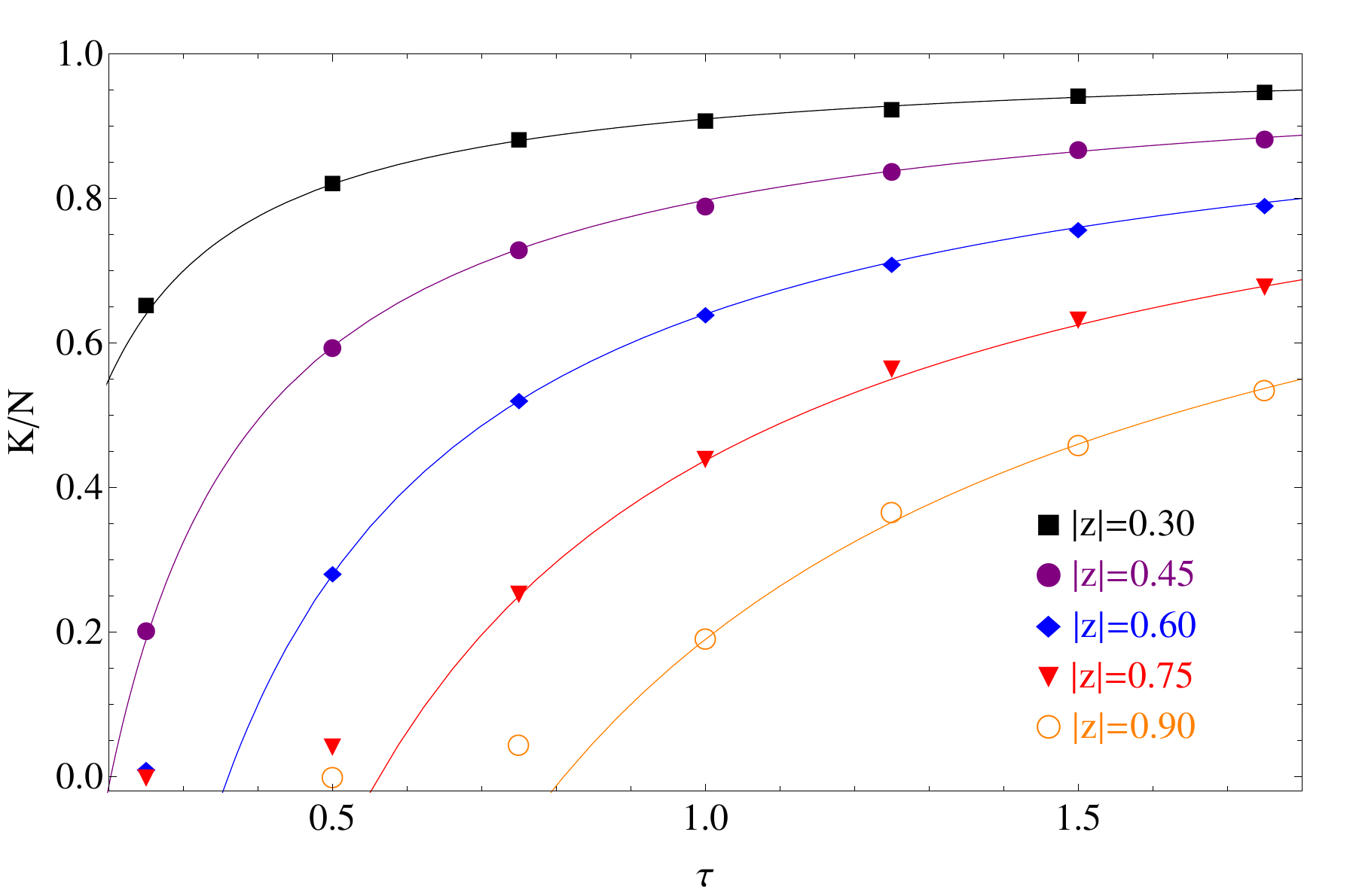}
\caption{The figure depicts theoretical (lines) and numerical (symbols) time dependence of the Petermann factor (rescaled by $1/N$), for different values of $|z|$. For the later, $3\cdot 10^4$, $200\times 200$ matrices were used. The discrepancy of the low lying points is a finite $N$ effect.
}
\label{plot2}
\end{figure}

Having an explicit solution for $v$ (\ref{mainv}), we can turn to $g$ (\ref{maing}).
Actually, one can show that $g$  also fulfills an exact for any $N$  Burgers-like equation
\be
\partial_{\tau} g = v \partial_r g + \frac{1}{N} \Delta_r g,
\label{twoBurgers1}
\ee 
which in the inviscid limit reduces to $\partial_{\tau} g = v \partial_z g$ or
\be
\partial_{\tau} g = 2 v \partial_z v,
\ee
if one uses $\partial_r g = 2 \partial_z v$ to eliminate $g$ from the 
right hand side. We see that we can calculate $g$ by differentiating $v$. 
The initial condition $X_0=0$ corresponds to 
$g_0(r)= \bar{z}/(|z|^2 + r^2)$, in particular $g_0(r=0)=1/z$.
For $v=0$ we have $\partial_{\tau}g=0$ so $g$ is constant in time, and
therefore it is equal to $g=1/z$ everywhere outside the shock line. 
Inside the shock line, we employ the second solution of  
(\ref{circle}), which via elementary integration leads to $g=\bar{z}/\tau + f(z)$.  Since both solutions have to match on the line of the shock due to condition (\ref{circle}), the arbitrary analytic  function $f$ has to be equal to zero. 
Note that for $r=0$, $g$ coincides with the electric field $G(z,\bar z)$ in the standard formulation mentioned earlier, 
so the average spectrum of the considered ensemble reads
\be
\rho(z,\tau)= \frac{1}{\pi \tau} \Theta(\tau-|z|^2),
\label{rhoz}
\ee
where $\Theta(x)$ is the Heaviside step function. We see that
complex eigenvalues are uniformly distributed on a growing disc of
radius $\sqrt{\tau}$. 

Finally, we would like to comment on the solution for large but finite $N$, at the vicinity of the shock. 
Since finite size implies non-zero viscosity, the dissipative term will regularize the shock leading to the smoothening of the sharp cliff of the eigenvalue density at the edge of the disk (\ref{rhoz}). Explicit calculations show that this is indeed the case. The smoothening makes the density at the edge to assume a universal shape given
by the complementary error function \cite{KS}. 
The argument goes as follows. We use the result of \cite{AKEMANN}, that the spectral density (diagonal part of the kernel) for the Ginibre ensemble is proportional to the $r \to 0$ limit of the characteristic determinant $D$ of the type considered here. The proportionality factor is the normalization $C_N$ and the Gaussian weight $p(z)= \exp (-\frac{N}{\tau} |z|^2)$, i.e. 
\be
\rho(z,\tau) \stackrel{N \to \infty}{=} C_N~ p(z) D(z,r \to 0,\tau),
\ee
with $C_N = \frac{2}{\tau \pi} \frac{1}{(N-1)!} \left ( \frac{N}{\tau} \right )^N$. 
Then, we may use the fact that the form of $D$ is exactly known for our initial conditions, since it represents the solution for the radial diffusion \cite{POLYANIN, BNWWISHART1, BNWWISHART2} 
\be
D=\int_0^{\infty} q e^{-N \frac{q^2+r^2}{\tau}} I_0 \left (\frac{2Nqr}{\tau} \right ) (q^2+|z|^2)^N dq.
\label{heat}
\ee 
A careful analysis of the saddle points shows that for large $N$ the main
contribution to the integral comes from quantities which scale asymptotically
as: $q =\theta  N^{-1/4}$, $|z|-\sqrt{\tau} =\eta N^{-1/2}$ and $r = \omega N^{-3/4}$,
for  $\theta$, $\eta$ and $\omega$ of order one.
We postpone details for a future publication. Here we note, however,  that this scaling 
is identical to the critical scaling for the cusp singularity of the Wishart/chiral random matrices. The reason for this lies in the functional form of the determinants, 
which happens to be identical for the two ensembles. In this way we establish additionally a somehow unexpected link between the universal scaling behavior for the Wishart and Ginibre ensembles. Taking first the large $N$ limit and then setting $\omega=0$, we recover from (\ref{heat}) a well-known result for the universal scaling at the spectral edge of the Ginibre ensemble
\be
\rho(\eta) \approx \frac{1}{2 \pi \tau} {\rm Erfc} \left ( \sqrt{\frac{2}{\tau}} \eta \right ).
\ee

We conclude this note with several remarks. 
First, it is inspiring to compare the Burgers-like structures even between the simplest hermitian model (GUE) and its non-hermitian counterpart, i.e. the Ginibre Ensemble. In the case of GUE, the characteristic determinant $D_{GUE}(z)$ fulfills a complex diffusion equation $\partial_{\tau}D_{GUE}=-\frac{1}{2N}\partial_{zz}D_{GUE}$.  The corresponding Burgers equation resulting from the Cole-Hopf transformation is complex too and has to be solved with complex characteristics. Singularities (shock waves) appear at discrete points (endpoints of the spectra) in the flow of eigenvalues~\cite{BN1}. On the contrary, for the GE, singularities are given by one-dimensional curves appearing in the flow of eigenvector correlations. The fact that in the Hermitian case the viscosity is formally negative also has far-reaching consequences. In particular, it is not smoothening the shock, like in the GE (where we observe the Erfc smearing), but it triggers violent oscillations, being the source of Airy universality. Similar universal oscillations originate from negative viscosity in other ensembles (complex Wishart (Laguerre) ensembles and complex Circular ensembles).   The fact that ensembles as different as GUE, CUE, LUE and GE have a similar underlying  mathematical structure of Burgers-like equations is remarkable and deserves further studies.

Moreover, for clearness we have only considered the dynamics of the simplest non-Hermitian ensemble, i.e. of the freely diffusing Ginibre matrices. Our approach works however for any initial condition imposed on the considered process. Additionally, the method can be used to study other non-Hermitian ensembles (e.g. non-Gaussian ones), for which the described coevolution will also be present. The resulting equations are of course much more involved in more general scenarios.
Our formalism could also be exploited to expand the area of application of non-Hermitian random matrix ensembles within problems of growth~\cite{LOEWNER}, charged droplets in quantum Hall effect ~\cite{HALL} and gauge theory/geometry relations in string theory~\cite{VAFA} beyond the subclass of complex matrices represented by normal matrices. 
  
Finally, we would like to emphasize, that a consistent description of non-Hermitian ensembles requires the knowledge of the detailed dynamics not only on the complex $z$ plane, where eigenvalues live, but also in the "orthogonal" $w$ plane.  In several standard techniques of non-Hermitian random matrix models this second variable is treated as an auxiliary parameter, serving as a regulator only.  We have shown that it governs, in the large $N$ limit,  the evolution of the standard correlator of eigenvectors which is furthermore coupled to the dynamics of the resolvent. Eigenvectors and eigenvalues evolve therefore simultaneously, and this coevolution is probably a common feature of all, also multi-point Green's functions in non-Hermitian random matrix models.


\section{Acknowledgments}
MAN appreciates discussion with Neil O'Connell on non-Hermitian Brownian walks, which triggered the interest in this problem. JG, MAN and PW would like to thank Jean-Paul Blaizot and Bertrand Eynard for fruitful conversations.
PW is supported by the International PhD Projects Programme of the Foundation
for Polish Science within the European Regional Development Fund of the European Union, agreement no. MPD/2009/6 
and the ETIUDA scholarship under the agreement no. UMO-2013/08/T/ST2/00105  of the National Centre of Science. 
MAN, ZB, WT and JG are supported by the Grant DEC-2011/02/A/ST1/00119 of the National Centre of Science. 




\begin{thebibliography}{99}
\bibitem{GG}
J. Ginibre, {\it J. Math. Phys.} {\bf 6}, 440 (1965).
\bibitem{KAROLZ}
W. Bruzda, V. Cappellini, H.-J. Sommers, K. \.Z{}yczkowski, {\it Physics Letters A} {\bf 373}, 320 (2009).
\bibitem{CHEMICALQCD}
H. Markum, R. Pullirsch, T. Wettig, {\it Phys. Rev. Lett.} {\bf 83}, 484 (1999).
\bibitem{BT} 
Ch. Biely, S. Thurner, {\it Quant. Finance} {\bf 8}, 705 (2008).
\bibitem{NEURAL}
H.-J. Sommers, A. Crisanti, H. Sompolinsky, Y. Stein, {\it Phys. Rev. Lett.} {\bf 60}, 1895 (1988).
\bibitem{CHALKERMEHLIG}
J.T. Chalker, B. Mehlig, {\it Phys. Rev. Lett.} {\bf 81}, 3367 (1998).
\bibitem{PETER1}
K. Frahm, H. Schomerus, M. Patra, C. W. J. Beenakker, {\it Europhys. Lett.} {\bf 49}, 48 (2000).
\bibitem{CHSCATTER}
Y. V. Fyodorov, B. Mehlig, {\it Phys. Rev. E} {\bf 66}, 045202 (2002).
\bibitem{HVH}
G. Hackenbroich, C. Viviescas, F. Haake, {\it Phys.  Rev. Lett.} {\bf 89}, 083902 (2002).
\bibitem{PETERMANN}
K. Petermann, {\it IEEE J. Quantum Electron.} {\bf 15}, 566 (1979).
\bibitem{DYSON}
F.J. Dyson, {\it J. Math. Phys.} {\bf 3}, 1191 (1962).
\bibitem{MAJUMDAR}
P. J. Forrester, S. N. Majumdar, G. Schehr, {\it Nucl. Phys. B} {\bf 844}, 500 (2011).
\bibitem{KATORI}
N. Kobayashi, M. Izumi, M. Katori, {\it Phys. Rev. E} {\bf 78}, 051102 (2008).
\bibitem{SCHEHR}
G. Schehr, S. N. Majumdar, A. Comtet, J. Randon-Furling, {\it Phys. Rev. Lett.} {\bf 101}, 150601 (2008).
\bibitem{LOEWNER}
R. Teodorescu, E. Bettelheim, O. Agam, A. Zabrodin, P. Wiegmann, {\it Nucl. Phys. B} {\bf 704}, 407 (2005).
\bibitem{NADAL}
C. Nadal, S. N. Majumdar, {\it Phys. Rev. E} {\bf 79}, 061117 (2009).
\bibitem{BN1}
J.-P. Blaizot, M.A. Nowak, {\it Phys. Rev. E} {\bf 82}, 051115 (2010).
\bibitem{BNW1}
J.-P. Blaizot, M.A. Nowak, P. Warcho\l{}, {\it Phys. Rev. E} {\bf 87}, 052134 (2013).
\bibitem{NEUBERGER}
H. Neuberger, {\it Phys. Lett. B} {\bf 670}, 235 (2008).
\bibitem{EWA}
E. Gudowska-Nowak, R. A. Janik, J. Jurkiewicz, M. A. Nowak, {\it Nucl. Phys. B} {\bf 670}, 479 (2003).
\bibitem{NEUBERGERPRL}
R. Lohmayer, H. Neuberger, {\it  Phys. Rev. Lett.} {\bf 108}, 061602 (2012) and references therein.
\bibitem{BN0}
J.-P. Blaizot, M. A. Nowak, {\it Phys. Rev. Lett.} {\bf 101}, 102001 (2008).
\bibitem{BNWWISHART2}
J.-P. Blaizot, M. A. Nowak, P. Warcho\l{}, {\it Phys. Lett. B} {\bf 724}, 170 (2013).
\bibitem{JANIKNOWAK}
R. A. Janik, M. A. Nowak, G. Papp, I. Zahed, {\it Nucl. Phys. B} {\bf 501}, 603 (1997).
\bibitem{JAROSZNOWAK}
A. Jarosz, M. A. Nowak, {\it J. Phys. A} {\bf 39}, 10107 (2006).
\bibitem{GIRKO}
V. L. Girko, {\it Theory Probab. Appl.} {\bf 29}, 694 (1983).
\bibitem{FEINBERGZEE}
J. Feinberg, A. Zee, {\it Nucl. Phys. B} {\bf 504}, 579 (1997).
\bibitem{CHALKERWANG}
J. T. Chalker, Z. J. Wang, {\it Phys. Rev. Lett.} {\bf 79}, 1797 (1997).
\bibitem{BJN} 
Z.Burda, R.A.Janik, M.A.Nowak, {\it Phys. Rev.} {\bf E 84}, 061125 (2011).
\bibitem{COLEHOPF}
J.D. Cole, {\it Quart. Appl. Math.} {\bf  9}, 225 (1951); 
E. Hopf, {\it Comm. Pure Appl. Math.} {\bf 3}, 201 (1950).
\bibitem{NOWAKNOER}
R. A. Janik, W. Noerenberg, M. A. Nowak, G. Papp, I. Zahed, {\it Phys. Rev. E} {\bf 60}, 2699 (1999).
\bibitem{FYODKHOR}
Y. V. Fyodorov, B. A. Khoruzhenko, {\it Commun. Math. Phys.} {\bf 273}, 561 (2007). 
\bibitem{KS} B. A. Khoruzhenko and H.-J. Sommers, “Non-Hermitian Random Matrix Ensembles”, chapter 18 in “The Oxford Handbook of Random Matrix Theory”, (Eds.) G. Akemann, J. Baik, P. Di Francesco, Oxford University Press, Oxford 2011.
\bibitem{AKEMANN}
G. Akemann, G. Vernizzi, {\it Nucl. Phys. B} {\bf 660}, 532 (2003).
\bibitem{POLYANIN}
A. D. Polyanin, V. F. Zaitsev, {\it Handbook of nonlinear partial differential equations} (Chapman \& Hall, London, 2004), p.31.
\bibitem{BNWWISHART1}
J.-P. Blaizot, M. A. Nowak, P. Warcho\l{}, {\it to be published in  Phys. Rev. E}, arXiv:1306.4014.
\bibitem{HALL}
O. Agam, E. Bettelheim, P. Wiegmann, A. Zabrodin, {\it Phys. Rev. Lett.} {\bf 88}, 236801 (2002).
\bibitem{VAFA}
V. A. Kazakov, A. Marshakov, {\it J. Phys. A} {\bf 36}, 3107 (2003).
\end{thebibliography}
\end{document}